\documentclass[pre,twocolumn,amsmath]{revtex4}
\usepackage{bm,natbib}
\newcommand{\fig}[3]{
\bigskip
\centerline{\epsfxsize=0.45\textwidth \epsfbox{#2}}
\smallskip {\small\noindent FIG. #1. #3} \bigskip }

\begin{document}
\input epsf

\title{Interface Unbinding in Structured Wedges}

\author{Gilberto Giugliarelli}
\affiliation{Dipartimento di Fisica, Universit\`a di Udine,
I--33100 Udine, Italy}

\affiliation{INFM--Dipartimento di Fisica, Universit\`a di Padova,
I--35131 Padova, Italy}


\begin{abstract}
The unbinding properties of an interface near structured wedges
are investigated by discrete models with short range interactions.
The calculations demonstrate that interface unbinding take place
in two stages: $i$) a continuous filling--like transition in the
pure wedge--like parts of the structure; $ii$) a conclusive
discontinuous unbinding. In 2$D$ an exact transfer matrix approach
allows to extract the whole interface phase diagram and the
precise mechanism at the basis of the phenomenon. The Metropolis
Monte Carlo simulations performed in 3$D$ reveal an analogous
behavior. The emerging scenario allows to shed new light onto the
problem of wetting of geometrically rough walls.
\end{abstract}


\maketitle

\section{Introduction}

Wetting phenomena concerns the properties of the liquid
film which forms when an undersaturated vapor (fluid) is put in
contact with a solid inert substrate. The critical properties of
the liquid--vapor interface are, generally, determined by nature
and range of the intervening interactions. In this respect, e.g.
for a planar substrate, different interactions potentials
determine the applicability of partial or complete wetting regimes
as well as the nature of the wetting transition (for a review on
these phenomena see \cite{dietrich1988, forgacs1991}).

However, substrate surface geometry can strongly influence these
properties. For example, in the complete wetting regime,
adsorption isotherms of random rough substrates \cite{pfeifer1989,
kardar1990} and linearly sculpted substrates \cite{parry2000,
rascon2000, bruschi2003} exhibit unusual exponents which are
determined by the surface geometry. On the other hand, while for
planar geometry partial wetting prevents the growth of macroscopic
films, in the same conditions, in pure wedges, we can have
continuous filling phenomena by which the film thickness is driven
to infinity \cite{parry1999, parry2000}. Finally, and this is
particularly relevant for wetting critical properties, there are
many indications that increasing surface roughness can change the
order of the wetting transition to first--order
\cite{giugliarelli1996, giugliarelli1997, stella1998,
giugliarelli1999}.

In this paper we report the results of an accurate investigation
about the unbinding properties of a thermally fluctuating
interface in wedge--modified systems (see Fig. 1), we denoted {\it
structured wedges} (SW). The choice of such a system structure is
motivated by the fact that they incorporate contrasting
geometrical motives (like wedges and ridges) which are typical of
the geometry of rough surfaces. In this respect, while the
separate effects of these geometries on wetting phenomena are
already rather well known \cite{parry1999,parry2003}, the effects
of their combination were never approached in detail before. Here,
we consider the study of both 2$D$ (Fig. 1$(a)$) and 3$D$ (Fig.
1$(b)$) SWs, and our results show that such structures implies two
stage interface unbinding transition, one of which discontinuous.
Our analysis of the phenomenon allows to understand its intimate
connection with the surface geometry as well as to make some
extension to the wetting properties of rough boundaries.

\section{The Model}

In the framework of solid on solid (SOS) approach our model
liquid--vapor interface corresponds to a lattice random walk (in
2$D$) or to a random surface (in 3$D$) in the vicinity of a fixed
substrate boundary. If $H_X$ denotes the substrate boundary
(integer) height at the position $X$ (depending on 2$D$ or 3$D$
space dimension, $X$ denotes a single variable $x$ or a couple of
variable $(x,y)$, respectively), the interface configurations can
be specified in terms of the local relative (integer) height
variables $z_X$ (see Fig. 2). The discrete nature of the model,
implies that non horizontal walls (see the magnifying glass in
Fig. 1) are shaped as staircases with slopes not exceeding unity.
Consequently, the height function $H_X$ satisfy the conditions
$H_{x+1}-H_x=0,\pm 1$ and $H_{x+1,y}-H_{x,y}=0,\pm 1$ in 2$D$ and
3$D$, respectively. In 3$D$, because of the translation invariance
along $y$ axis, one has also $H_{x,y+1}-H_{x,y}=0$.

\fig{1}{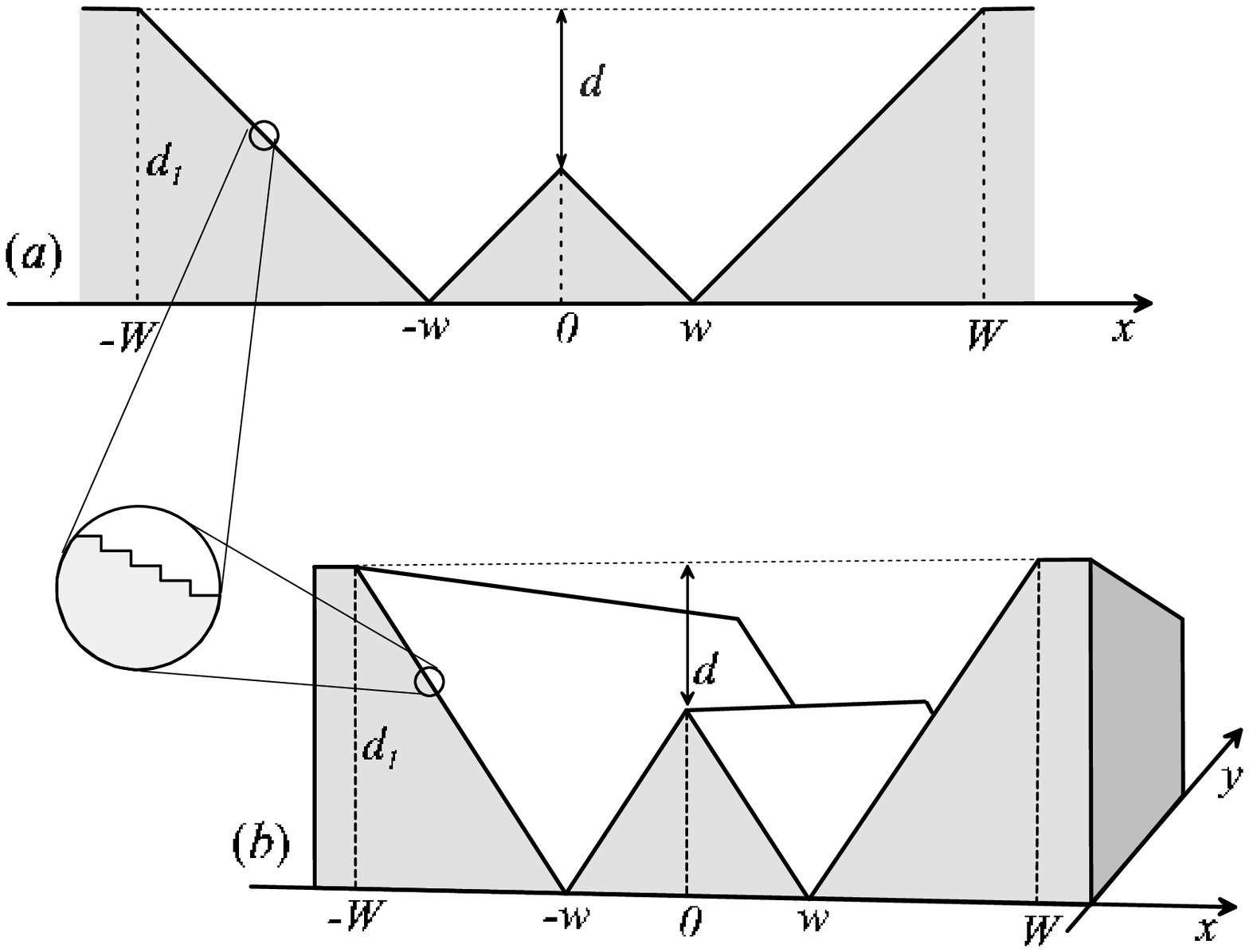}{Sketch of the 2$D$ (panel $(a)$) and 3$D$
(panel $(b)$) SW geometries studied in this paper. The {\it
magnifying glass} shows the staircase nature of the tilted smooth
walls (with slope $|1/n|$) by which the SW are constructed.}

At coexistence (no chemical potential differences with respect to
its critical bulk value), an interface (see Fig. 2$(a),(b)$), can
be studied in terms of an Hamiltonian of the form
\begin{equation}\label{Hamiltonian} \mathcal{H}=
\sum_{\langle X,X'\rangle}
\mathcal{E}(1+|h_{X}-h_{X'}|^\gamma)-\mathcal{U}
\sum_{X}\delta_{z_X,0},
\end{equation}
where the first sum is done over all pairs of nearest neighbor
columns. In the above expression $\mathcal{E}$ and $-\mathcal{U}$
(with $\mathcal{E},\mathcal{U}>0$) are the energy cost of any
interface step (or plaquette in 3$D$) and the energy gain of each
interface contact (of horizontal step or plaquette) with the
substrate, respectively; $\gamma=1$ or $\infty$ determines the SOS
or restricted SOS (RSOS) character of the implemented walk model,
respectively. To limit the computational complexity, in 3$D$ we
consider only RSOS interfaces.

\fig{2}{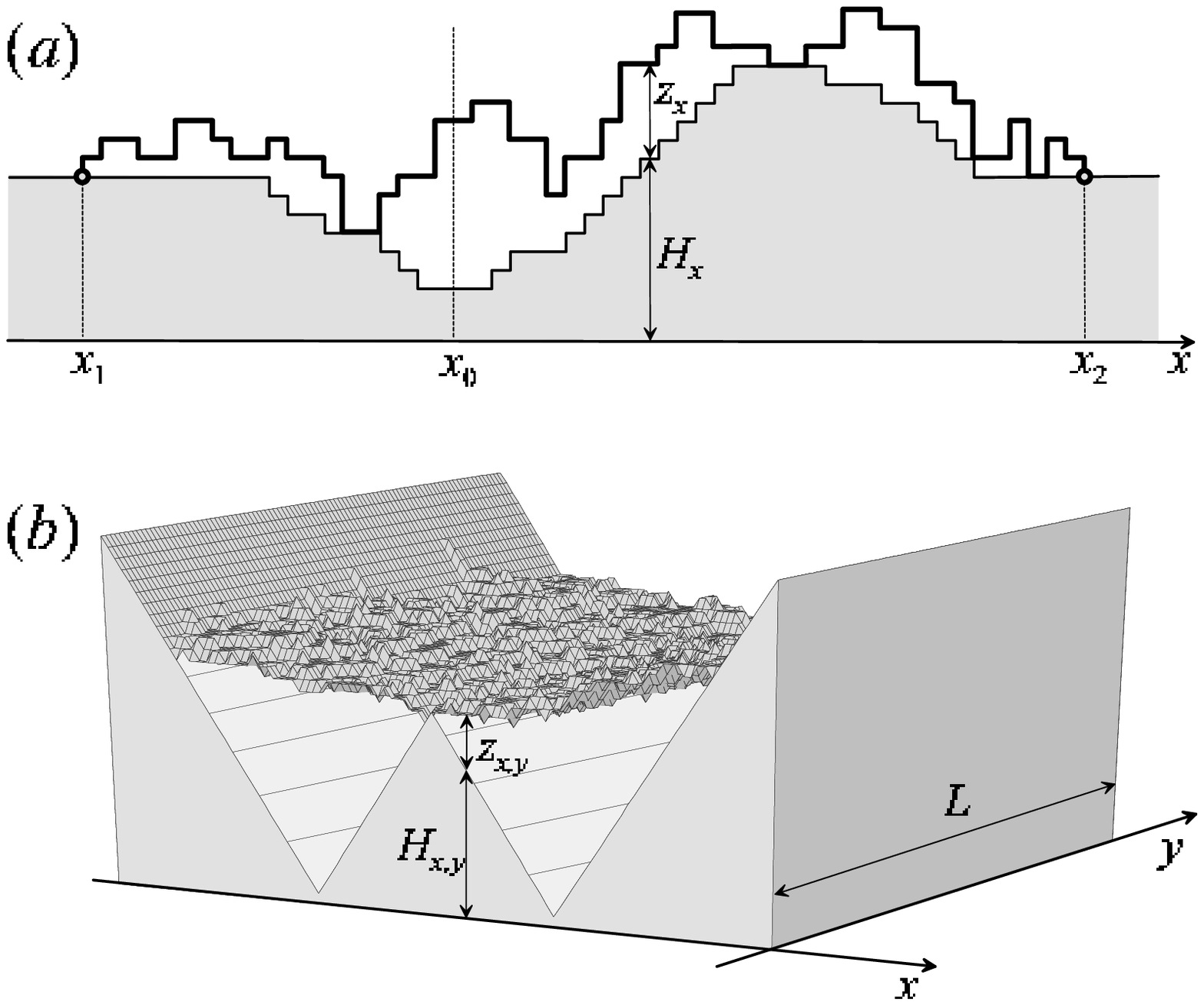}{$(a)$ Example of a configuration of a
directed walk (heavy continuous line) with the ends anchored to a
2$D$ boundary (light continuous line delimiting the shaded area).
$(b)$ An RSOS interface--surface configuration near our 3$D$ SW.}

\section{2$D$ Structured Wedges}

In 2$D$, an interface in the vicinity of a boundary (with a
generic shape) can be fully treated by a transfer matrix approach
based on the matrices $\hat{\bm{R}}_{x}$ and $\hat{\bm{L}}_{x}$
defined as follows
\begin{subequations}\label{Rx-Lx}
\begin{eqnarray}
{[\hat{\bm{R}}_x]_{z,z'}} &=& {\omega^{|z'-z+H_{x+1}-H_x|^\gamma}
k^{\delta_{z',0}}},\label{R_x}
\\
{[\hat{\bm{L}}_x]_{z,z'}} &=& {\omega^{|z'-z+H_{x-1}-H_x|^\gamma}
k^{\delta_{z',0}}}.\label{L_x}
\end{eqnarray}
\end{subequations}
Here $\omega= \hbox{e}^ {-\mathcal{E}/k_BT}= \hbox{e}^{-1/t}$
($t=k_BT/\mathcal{E}$) and $k=\hbox{e}^{U/k_BT}=\hbox{e}^{u/t}$
($u=U/\mathcal{E}$) correspond to step and wall fugacities and, by
definition, $\omega [R_x]_{z,z'}$ and $\omega [L_x]_{z,z'}$ are
the Boltzmann weights of the elementary walks from $(x,H_x+z)$ to
$(x+1,H_{x+1}+z')$ or $(x-1,H_{x-1}+z')$, respectively. Thus, the
partition function of, e.g. the walks between $(x_1,H_{x_1})$ and
$(x_2,H_{x_2})$ (see Fig. 2$(a)$), is given by
\begin{equation}\label{ZR-ZL}
\mathcal{Z}_{x_1,x_2} = \omega^{x_2-x_1} [\hat{\bm{R}}_{x_1\to x_2
}]_{0,0} \equiv \omega^{x_2-x_1}[\hat{\bm{L}}_{x_1\leftarrow x_2
}]_{0,0},
\end{equation}
with $\hat{\bm{R}}_{x\to x'}\equiv \prod_{i=x}^{x'-1}
\hat{\bm{R}}_{i}$ and $\hat{\bm{L}}_{x\leftarrow x'}\equiv
\prod_{i=x'}^{x+1} \hat{\bm{L}}_{i}$

But the transfer matrix approach allows also the direct
calculation of the distance probability distribution function
(PDF) at any position along the boundary. A walk like the one in
Fig. 2$(a)$ can be divided in two parts, e.g. at $x_0$, obtaining
two walks with Boltzmann weights $[\hat{\bm{R}}_{x_1\to
x_0}]_{0,z}$ and $[\hat{\bm{L}}_{x_0\leftarrow x_2}]_{0,z}$,
respectively. Thus, the walk distance PDF at $x_0$ is proportional
to the product $[\hat{\bm{R}}_{x_1\to x_0}]_{0,z}\cdot
[\hat{\bm{L}}_{x_0\leftarrow x_2}]_{0,z}$. On the other hand,
\emph{storing} the initial walk distance distributions, at $x_1$
and $x_2$, in the vectors $\bm{r}_{x_1}$ and $\bm{l}_{x_2}$ (in
the case of Fig. 2$(a)$ we should set $[\bm{r}_{x_1}]_z=
[\bm{l}_{x_2}]_z= \delta_{z,0}$), the following iterations
\begin{equation}\label{r_l_propagation}
\bm{r}_{x+1}= \bm{r}_{x} \hat{\bm{R}}_x;\ \ \ \ \ \ \ \ \
\bm{l}_{x-1}= \bm{l}_{x} \hat{\bm{L}}_x,
\end{equation}
allows to get the quantities $[\hat{\bm{R}}_{x_1\to x_0}]_{0,z}$
and $[\hat{\bm{L}}_{x_0\leftarrow x_2}]_{0,z}$ in terms of the
components of $\bm{r}_{x_0}$ and $\bm{l}_{x_0}$ vectors. Indeed,
iterations (\ref{r_l_propagation}) transfer toward right and left
the local walk distance distributions to any position along the
boundary, like in a forward diffusion process. Therefore, in the
limit of an indefinite walk (our interface model), i.e.
$|x_0-x_{1,2}|\to \infty$, the transferred right and left PDFs,
defined as
\begin{equation}\label{r_l_profiles}
P_{x_0}^{(R)}(z)={1\over{[\bm{r}_{x_0}]}}[\bm{r}_{x_0}]_z;\ \ \ \
\ \ \ P_{x_0}^{(L)}(z)={1\over{[\bm{l}_{x_0}]}}[\bm{l}_{x_0}]_z,
\end{equation}
will reach steady profiles depending only on the specific boundary
conformation. In these conditions the interface PDF at any
position $x_0$ will be given by
\begin{equation}\label{PDF_x0}
P_{x_0}(z)= P_{x_0}^{(R)}(z)P_{x_0}^{(L)}(z),
\end{equation}
and the quantity
\begin{equation}\label{fe_profile}
\Delta f_{x_0} = -\ln P_{x_0}(z)=-\left[\ln P_{x_0}^{(R)}(z)+\ln
P_{x_0}^{(L)}(z)\right],
\end{equation}
can be seen as the corresponding local excess interface free
energy profile.

Our SWs boundaries contain only unit vertical steps. Thus, the
transfer matrices (\ref{Rx-Lx}) can be of only three different
forms, i.e. those corresponding to a down step, an up step and no
vertical step; we denote these matrices as ${\hat{\bm{D}}}$,
${\hat{\bm{U}}}$ and ${\hat{\bm{F}}}$, respectively. For tilted
wall regions with slope $|1/n|$, the calculation of right and left
PDFs can be done by implementing iterations
(\ref{r_l_propagation}) in terms of
$\hat{\bm{F}}^{n-1}\hat{\bm{D}}$ or
$\hat{\bm{F}}^{n-1}\hat{\bm{U}}$ matrices (depending on the
descending or ascending character of the wall); for large system
sizes, the corresponding $P^{(R)}$ and $P^{(L)}$ will
asymptotically become eigenvectors of these matrices. In
particular, in binding conditions (i.e. at high enough $u$
values), the interface PDFs at $x=\pm w$ and $x=0$ (see Fig.
1$(a)$) will correspond to $\bm{\Phi}_{n}^2$ and
$\bm{\Psi}_{n}^2$, $\bm{\Phi}_{n}$ and $\bm{\Psi}_{n}$ being the
bound eigenvectors of matrices $\hat{\bm{F}}^{n-1}\hat{\bm{D}}$
and $\hat{\bm{F}}^{n-1}\hat{\bm{U}}$, respectively. In this
respect, an accurate investigation allows the determination of
these states (exact for $n=1,2$, numerical for $n>2$) both for SOS
as well as RSOS walk models.

A brief analysis of $n=1$ case (tilted walls with unit slope) is
particularly useful for the understanding of the general scenario.
Matrices $\hat{\bm{F}}^{n-1}\hat{\bm{D}}$ and
$\hat{\bm{F}}^{n-1}\hat{\bm{U}}$ reduce to $\hat{\bm{D}}$ and
$\hat{\bm{U}}$ and for SOS walks both $\bm{\Phi}_1$ and
$\bm{\Psi}_1$ components scale as $\mu_{D_1}^z$ and $\mu_{U_1}^z$,
respectively. $\mu_{D_1}={{1+k\omega^2}\over{k\omega}}$, and thus
the local character of $\bm{\Phi}_1$ (i.e. $\mu_{D_1}<1$) is
achieved until $k>k_{D_1,c}$, with
$k_{D_1,c}={1\over{\omega(1-\omega)}}$. The corresponding
interface PDF at the SW bottoms (i.e. $x=\pm w$) is $P_{\pm
w}(z)=\mu_{D_1}^{2z}$ and the resulting interface average
distance, $\langle z\rangle_{\pm w}= {{\sum_{z} z\cdot P_{\pm
w}(z)}\over{\sum_{z} P_{\pm w}(z)}}$ is given by
\begin{equation}\label{}
\langle z\rangle_{\pm w}=
{k\over{(2-k+k\omega^2)(1+2k\omega^2+k^2\omega^4-k^2\omega^2)}}.
\end{equation}
As $k\to k_{D_1,c}^+$, $\langle z\rangle_{\pm w}$ diverges with
continuity as $\left[k-k_{D_1,c}\right]^{-1}$ and thus $k_{D_1,c}$
is the critical fugacity associated to the continuous interface
unbinding from SW bottoms. This transition is the analog of the
critical filling transition in pure wedge geometry
\cite{parry2000,parry2002}. We have verified that for
$n=2,3,\ldots, 10$ one observe the same behavior (also for RSOS
walk model) and in Fig. 3 we show the corresponding phase--line
boundaries, consisting in the the plot of the quantity
$u_c(t)=t\ln k_{D_n,c}$. The picture demonstrate that in the wedge
geometry, in contrast to the case of flat substrates (see the
bottom curves in Fig. 3), a bound interface always (also at low
$t$) requires a finite attraction energy $u$. On the other hand,
the local minima in the phase--line boundaries confirm the
possibility of reentrance phenomena which were already predicted
\cite{giugliarelli1997} for rough self--affine boundaries. More
details about these aspects will be reported elsewhere
\cite{giugliarelli2004}.

\fig{3}{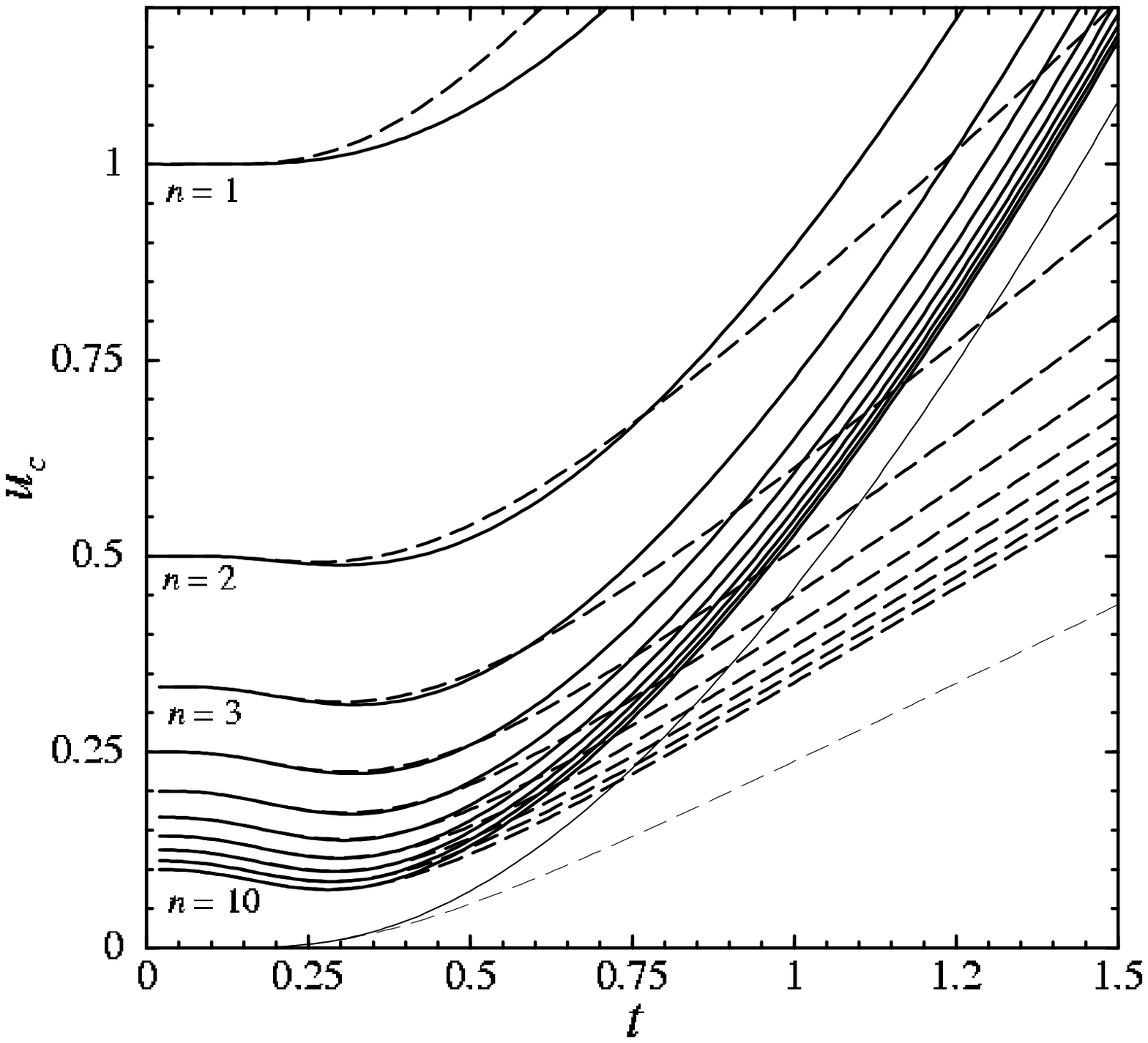}{SOS (continuous lines) and RSOS (dashed
lines) interface phase--line boundaries in the $u$--$t$ plane for
2$D$ SWs (like in Fig. 1$(a)$) with tilted wall slope $|1/n|$
($n=1,2,\ldots,10$). Upper curves ($n=1,2$) are the result of
exact calculations; the other curves were obtained by a numerical
implementation of iterations (\ref{r_l_propagation}). Lower
(light) curves are the interface phase--line boundary for a flat
wall \cite{abraham1980}.}

On the contrary, $\mu_{U_1}={{k\omega}\over{k-1}}$ and therefore
the local character of $\bm{\Psi}_1$ is achieved for $k>k_{U_1,c}$
with $k_{U_1,c}=1/(1-\omega)<k_{D_1,c}$. Thus, when $k$ approaches
$k_{D_1,c}$ from above, and the interface unbinds from SW bottoms,
it remains tightly bound to the central ridge. To be more precise,
in the limit $k\to [k_{D_1,c}]^+$ the mean interface distance from
the SW central ridge is finite and is given by
\begin{equation}\label{mean-Z-W}
\langle z\rangle_{0}=
{{\omega^4(1-\omega+\omega^2)}\over{(1+\omega+\omega^2)
(1+\omega^2)(1-\omega)^2}}.
\end{equation}

Finally, for $k<k_{D_1,c}$ we are out of binding conditions for
$\hat{\bm{D}}$ matrix and the lateral descending walls of the SW
are no more able to bind the walks: iterations
(\ref{r_l_propagation}), asymptotically produce delocalized right
and left PDFs. In these conditions, the presence of the central
ridge can have only very marginal effects (vanishing with the size
of the structure) and, thus, the interface unbinds from the SW.
The conclusion is: $k_{D_1,c}$ is also the threshold of a
discontinuous unbinding transition! The curves in Fig. 3 can be
seen also as the interface phase--lines for 2$D$ SW first--order
unbinding.

\fig{4}{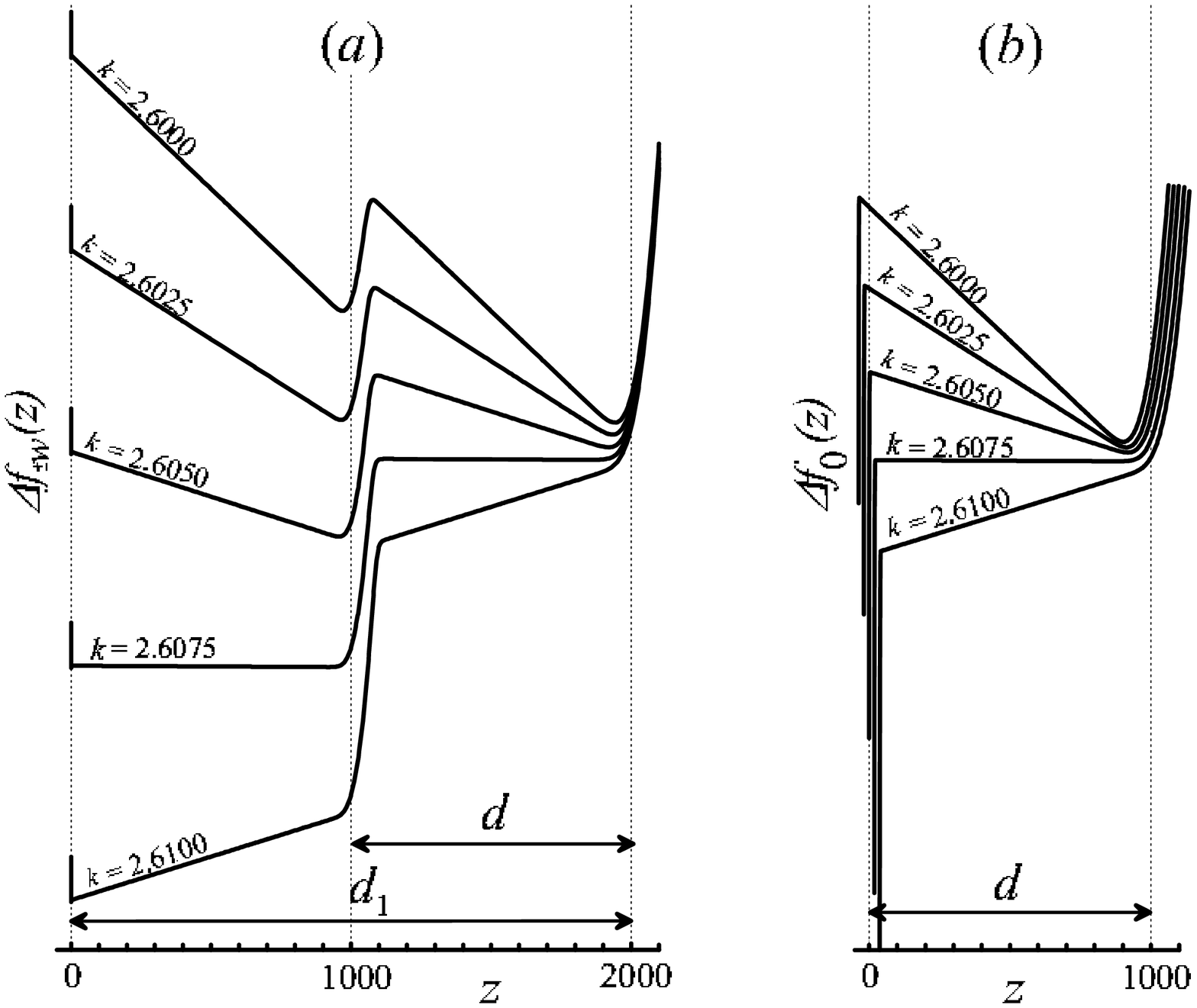}{Excess free energy profiles (in arbitrary
units) at $x=\pm w$ (panel $(a)$) and $x=0$ (panel $(b)$) for a
2$D$ SW with wall slope $|1/2|$, $d=1000$ and $d_1=2000$ (see Fig.
1$(a)$). The free energy profiles have been obtained by a
numerical implementation of iterations (\ref{r_l_propagation}) at
$\omega=0.2$ ($t=0.6213\ldots$). The curves are shifted vertically
to increase the picture clearness; in panel $(b)$ also a little
horizontal shift has been adopted to avoid curves superposition at
$z=0$.}

The conclusive validation of the above, surprising, conclusion is
obtained by the analysis of the free energy profiles. In Fig. 4 we
show a plot of $\Delta f_{\pm w}(z)$ (panel $(a)$) and $\Delta
f_{0}(z)$ (panel $(b)$) profiles obtained at fixed $t$ for an RSOS
interface in a finite size SW with wall slope $1/2$ (see the
specific SW dimensions in the figure legend) characterized by an
unbinding critical fugacity $k_{D_2,c}=2.607536\ldots$ (exact
calculation). Curves in panel $(a)$ show, as $k$ decreases, the
following free energy profile evolution: $i$) a unique free energy
minimum at $z=0$ (for $k=2.6100>k_{D_2,c}$) which delocalizes into
a wide one at $k=2.6075\simeq k_{D_2,c}^{(\text{RSOS})}$ (i.e. the
continuous transition in the pure wedges); $ii$) double minima
profiles for $k=2.6050, 2.6025 \lesssim k_{D_2,c}$ with the minima
placed at $z=d_1-d$ and $z=d_1$ (i.e. coexistence between the
state localized at the SW central ridge height and the bulk
unbound state); $iii$) dominance of bulk unbound state ($k=2.6000
< k_{D_2,c}$). An analogous  evolution is extracted by the free
energy profiles of $\Delta f_{0}$ in panel $(b)$. Therefore, in
passing between $ii$) and $iii$) we have a discontinuous
transition! Finally, the picture clarifies that the discontinuous
nature of such transition is strictly related to the fact that,
for diverging system size, $d$ parameter maintains larger than the
widths of the two free energy minima (scaling as $\sim W^{1/2}$).

\fig{5}{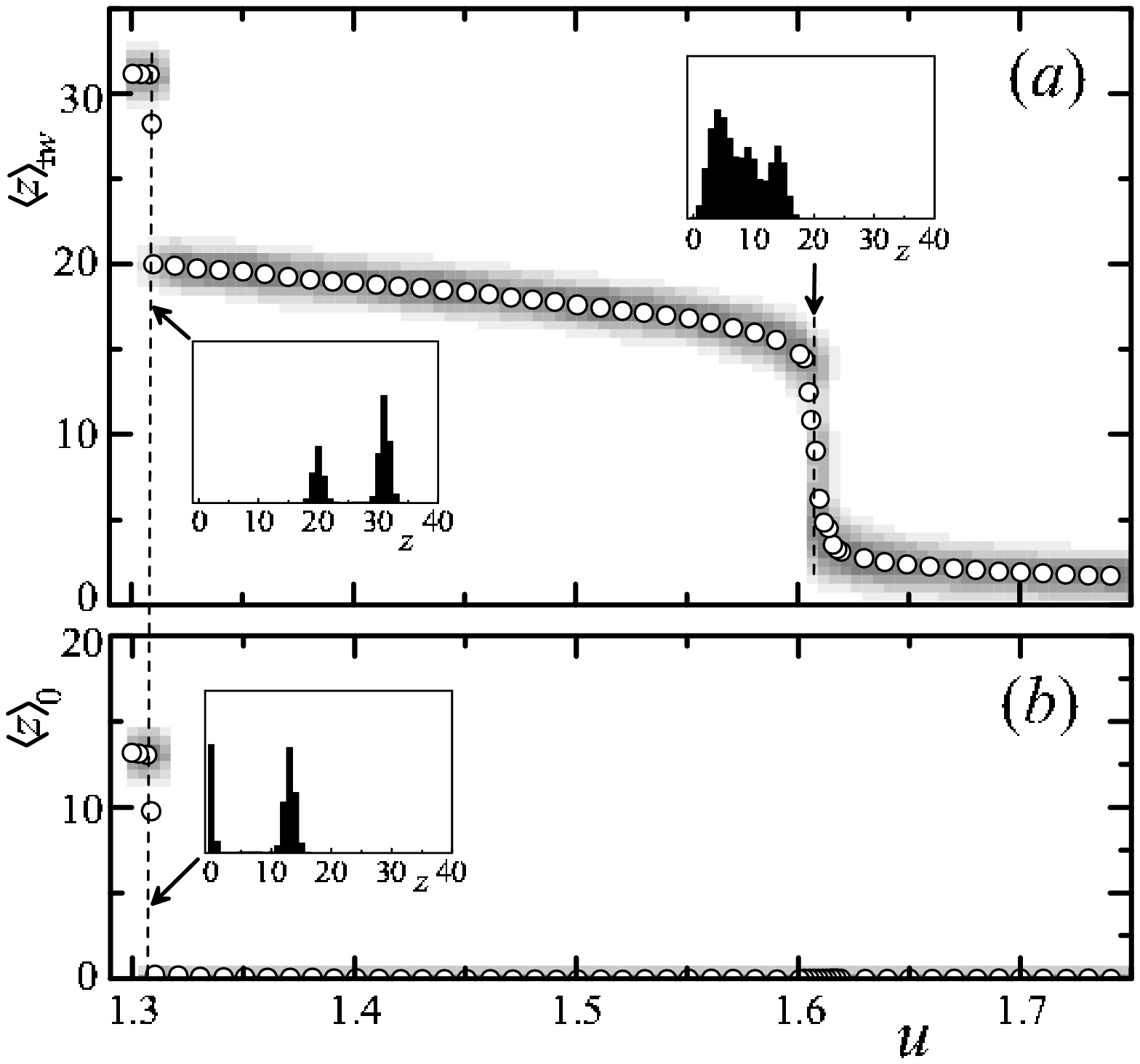}{Average interface distances (open dots)
from 3$D$ SW bottoms (panel $(a)$) and central ridge (panel $(b)$)
and the corresponding distributions (gray shaded area). The data
come from Metropolis Monte Carlo simulations at constant $t=2.0$
and system sizes $L=100$, $W=50$, wall unit slope and $d_1/d=2$
(i.e. $d=16$). In the insets is the detail of interface distance
distributions at the $u$ values marked by the dashed lines.}

\section{3$D$ Structured Wedges}

Transfer matrix approach is not applicable to the wetting problem
in 3$D$. In this case we have considered Metropolis Monte Carlo
simulations of an RSOS fluctuating interface (see Fig. 2$(b)$) in
a 3$D$ SW geometry like the one sketched in Fig. 1$(b)$. Because
of rapid increase of computation times with system size, we have
considered relatively small systems: the simulations were done for
squared system with size $L=50,100,150$ (in the $y$ axis
direction) and $W=L/2$ ($L$ and $W$ are in lattice units) and
using periodical boundary conditions along both $x$ and $y$ axes.
The SWs considered in the simulations were usually constructed by
(staircase) tilted walls with slope 1 or 1/2 and $d_1/d\simeq 2$
(see Fig. 1$(b)$). As usual, the implemented Metropolis Monte
Carlo algorithm was based on local moves corresponding to the
attempt of changing (at random) the local surface height by one
unit (i.e. $z_{x,y} \to z_{x,y}\pm 1$). After equilibration, the
calculation of equilibrium average parameters has been performed
on the basis of very long simulations of up to $10^5\div 10^6$
$MCS$ (1 $MCS \equiv L^2$ Monte Carlo moves).

In Fig. 5, as a representative of the general behavior, we show
the $u$ dependence of the interface averages distances from SW
bottoms ($\langle z\rangle_{\pm w}$, panel $(a)$) and SW central
ridge ($\langle z\rangle_{0}$, panel $(b)$) obtained a constant
$t$ for a given 3$D$ SW (see geometrical specifications in figure
legend). The behavior of these two quantities follows a scheme
very similar to the one outlined for 2$D$ SW interface unbinding:
$i$) a continuous filling--like transition of the two component
wedges at $u\simeq 1.607$ (note also in the right inset in panel
$(a)$ the roughly flat height distribution); $ii$) a discontinuous
detachment from the SW central ridge at $u\simeq 1.309$ (see the
double peak structure of the interface height distributions in the
left insets in panel $(a)$ and $(b)$). However, at least for the
finite size system considered here, continuous and discontinuous
unbinding seem to be separated by a finite $u$ gap. Most probably
this is due to the linear extension of the system along $y$ axis
which corresponds to the main difference between the present 3$D$
SW geometry and the 2$D$ one. With an extensive simulations
program, we planned to evaluate the whole interface phase--lines
for 3$D$ wedges with the intent to clarify also this aspect.

\section{Conclusions}

Our investigation demonstrate that interface unbinding in SWs
takes place in two stages: the first one corresponds to the
continuous filling transition of the two component wedges; in the
second stage the interface jumps, discontinuously, from a state
localized at the SW central ridge and the bulk unbound state. We
stress the fact that in SWs the bulk unbound state come into
field, rather than in some artificial way
\cite{nieuwenhuizen1988}, just because of the {\it corrugated}
geometry of the substrate. First--order nature of the final
unbinding transition is due to the {\it competition} between
binding at the central ridge and whole SW filling. In other words,
it is the spatial combination of wedges and ridges of SWs to
create the conditions for a discontinuous interface unbinding. As
seen in section III, a single ridge is only able to strongly pin
the interface retarding (with respect to a wedge) its unbinding
from the apex (for more details see also ref.
\cite{giugliarelli2004}). In complete wetting regime, as shown in
ref. \cite{parry2003}, interface unbinding near the apex mimics
planar critical wetting.

Generalization of our conclusions to more complex SWs and/or to
random rough surfaces should take into account, carefully, the
type and the scaling properties of the corrugation introduced by
the specific boundary. On the other hand, our analysis about
interface unbinding in 2$D$ SWs give some more efforts on the key
role of surface roughening exponent $\zeta_S$ in determining the
nature of the wetting transition in self--affine rough substrates.
In the light of the present results, we have started new
investigations of 2$D$ and 3$D$ rough geometries, with the purpose
to get conclusive insights in this issues.

\acknowledgments

The work has been partly supported by INFM.


\end{document}